\documentclass[a4paper,12pt,fleqn,reqno]{article}
\usepackage{amsmath,amsfonts,amssymb,graphicx, hyperref,epstopdf, amsthm,caption,enumerate ,a4}
\usepackage[numbers]{natbib}
\theoremstyle{plain}    \theoremstyle{remark}
  
\usepackage{titlesec,soul}
\titleformat{\section}[block]{\large\bfseries\filcenter}{\thesection.}{1em}{} \titlespacing*{\section}{0pt}{25pt}{5pt}
\numberwithin{equation}{section}\mathindent2em
\begin{document}

\begin{center}
\textbf{ Similarity solution for the flow behind a magnetogasdynamic exponential shock wave in a perfect gas with varying density, heat conduction and radiation heat flux}\\[.75em]
\textbf{Ruchi Bajargaan$^{a,1}$ \ , \  Arvind Patel$^{a,2}$\ and \ Manoj Singh$^{a,3}$}\\[.5em]
$^{a}$Department of Mathematics,\\
 University of Delhi, Delhi 110 007, INDIA\\
$^{1}$e-mail: duruchi11@gmail.com\\[.5em]
$^{2}$e-mail: apatel@maths.du.ac.in\\[.5em]
$^{3}$e-mail: smanojs2du@gmail.com
\end{center}

\vskip 1em\baselineskip16pt \noindent{\bf Abstract:}
Similarity solutions are obtained for one dimensional, unsteady, adiabatic propagation of an exponential shock wave in a perfect gas with heat conduction and radiation heat flux, in the presence of azimuthal magnetic field. The shock wave is driven out by a piston moving with time according to an exponential law. The equilibrium flow conditions are maintained. The heat conduction is expressed in terms of Fourier's law and the radiation is considered to be of the diffusion type for an optically thick grey gas model. The thermal conductivity and the absorption coefficient are assumed to vary with temperature and density according to power law. The density and magnetic field ahead of the shock front, are assumed to vary as an exponential law. The effects of the variation of the strength of ambient magnetic field, heat transfer parameters, adiabatic exponent, ambient density variation index on the shock strength, the distance between the piston and the shock front, and on the flow variables are studied out in detail. The similarity solution exists only when the sum of shock radius and ambient magnetic field exponent is equal to the half of the ambient density exponent. It is manifested that the shock strength decreases by increasing the strength of ambient magnetic field but it is independent from the heat transfer parameters. The total energy of the flow field behind the shock front is not constant but varies as power of shock radius. The compressibility of the medium is increased in the non-magnetic field. Also, the presence of the magnetic field have significant effects on the shock wave.


\noindent \textbf{Key Words:} Exponential shock wave; Self similar solution; Magnetic field; Conduction and radiation heat flux.

\section{Introduction}
Shock processes can normally take place in various astrophysical situations such as stellar winds, photo-ionized gas, supernova explosions, collisions between high velocity clumps of interstellar gas etc. Shock waves have tremendous importance in astrophysics, geophysics, plasma physics, nuclear science and interstellar masses for non linear systems and in several other fields. Shock waves are also associated with spiral density waves, quasars and radio galaxies. These are ordinary in the interstellar medium because of a great diversity of energetic events and supersonic motions, such as bipolar out flow from young protostellar objects, cloud-cloud collision, powerful mass losses by massive stars in a late stage of their evolution, central part of star burst galaxies, supernova explosions, etc. Same phenomena also takes place in laboratory situations, for example, when a piston is driven rapidly into a shock tube when a projectile moves supersonically through the atmosphere, in the blast wave produced by a strong explosion, or when rapidly owing gas come across a constriction in a flow channel or runs into a wall (Nath and Vishwakarma \cite{gv}). Shocks are present throughout the observed universe and are thought to play a critical role in the transportation of energy into the interstellar medium and setting motion processes observed in nebulae that finally could lead to the formation of stars. 

Radiation has a significant role in the several hydrodynamic processes related to shock waves and explosions because it plays an important role in energy transport over the vast distances encountered between stellar objects and, can significantly change the dynamics of a shock or blast wave. The role of radiation is not only limited to the luminescence of the heated body but it also disturbs the hydrodynamic movement of matter in the form of heat exchange and energy transfer.  Marshak \cite{marshak} has obtained similarity solutions of the radiation hydrodynamic equations for particular cases when there is planer symmetry, and radiation pressure and energy are negligible, although flux is important. Marshak \cite{marshak} considered the cases of (1) constant density, (2) constant pressure, and (3) power law time dependence of temperature. Elliot \cite{Elliot} considered the conditions holding self similarity with a described functional form of the mean free path of radiation and obtained a self similar solution for spherical explosions. Wang \cite{wang} discussed the problems of radiating walls, either moving or stationary, developing shocks at the head of self-similar flow field. Recently, the study of self similar solution of a shock wave in an ideal gas, non-ideal gas or dusty gas with the heat conduction and radiation heat flux has been done by many authors \cite{gv}, \cite{ru}, \cite{sahu2}, \cite{baj}, \cite{vishu}, \cite{kk}, \cite{vis}, \cite{vis11}.  

Magnetic fields spread throughout the universe and have critical roles in several astrophysical problems. All astrophysical plasmas are affected by magnetic field. Magnetic fields have a significant role in energy and momentum transport and can quickly release energy in flares. Magnetic fields are involved in many interesting problems. The shock waves in the presence of magnetic field in conducting perfect gas are important for interpretation of shocks in supernova explosion and explosion in the ionosphere. Complex filamentary structures in molecular clouds, shapes and the shaping of planetary nebulae, synchrotron radiation from supernova remnants, magnetized stellar winds, galactic winds, dynamo effects in stars, gamma-ray bursts, galaxies and galaxy clusters as well as other interesting problems all involve magnetic fields. The industrial applications are drag reduction in duct flows, control of turbulence of immersed jets in the steel casting process and advanced propulsion and flow control schemes for hypersonic vehicles, design of efficient coolant blankets in tokamak fusion reactors, involving applied external magnetic fields (see, Hartmann \cite{hart}, Balick \cite{bali} ).

The limiting case of a self-similar flow-field with a power law shock is the flow-field formed with an exponential shock indicated by Sedov \cite{Sedov} (see, Ranga Rao and Ramana \cite{rangarao}, Singh and Srivastava \cite{srivastava}, Vishwakarma and Nath \cite{vis1}, Vishwakarma and Nath \cite{vis2}). Ranga Rao and Ramana \cite{rangarao} have obtained approximate analytic solutions for unsteady self-similar motion of a perfect gas displaced by a piston according to an exponential law. Singh and Srivastava \cite{srivastava} have obtained the self-similar solution for flows of a perfect gas behind the cylindrical shock wave propagating exponentially in an atmosphere whose density varies inversely as the fourth power of shock radius. Vishwakarma and Nath \cite{vis1} have obtained the similarity solutions for the problem of unsteady self similar motion of exponential shock wave in a  dusty gas (a mixture of perfect gas and small solid particles). Vishwakarma and Nath \cite{vis2} have obtained the similarity solutions for the unsteady flow of non-ideal gas behind a strong exponential shock driven out by a piston. Singh et al. \cite{sin} have obtained the similarity solution for the unsteady flow of non-ideal gas behind an exponential shock wave with the effect of magnetic field . Nath \cite{na1} has obtained similarity solutions for one dimensional unsteady isothermal and adiabatic flows behind a strong exponential shock wave in a rotating, axisymmetric non-ideal gas by taking into account the variable azimuthal and axial fluid velocities. Nath and Sahu \cite{sahu1} have obtained a similarity solution behind an exponential shock wave in a rotational axisymmetric perfect gas with magnetic field by taking the variable density, the azimuthal and axial fluid velocities. Nath and Sahu \cite{sahu2} have obtained a self similar solution for an exponential shock wave in a rotating axisymmetric non-ideal gas with conduction and radiation heat flux. Nath and Singh \cite{sumeeta} have obtained self similar solution behind magnetogasdynamic exponential shock wave in a self-gravitating gas. Bajargaan and Patel \cite{baj} have obtained the similarity solution for the flow behind an exponential shock wave in a self-gravitating, rotating, axisymmetric dusty gas with heat conduction and radiation heat flux by assuming the variable azimuthal and axial fluid velocities.  In all of these works, study of self similar solution of the flow behind a magnetogasdynamic exponential shock wave in a perfect gas under the effect of heat conduction and radiation heat flux together with variable density has not been done. The present work is the extension to the work of Ranga Rao and Ramana \cite{rangarao} by taking conduction and radiation heat flux, azimuthal magnetic field and variable density into account. 

The purpose of this study is to obtain self similar solutions for the propagation of an exponential shock wave which is driven out by a piston or explosion moving with time according to an exponential law in a perfect gas with the effect of azimuthal magnetic field, heat conduction and radiation heat flux and variable density. The density ahead of the shock front is assumed to be decreasing and the azimuthal magnetic field ahead of the shock front is assumed to be increasing, constant and decreasing according to an exponential law. The equilibrium flow conditions are assumed to be maintained. Radiation pressure and radiation energy are assumed to be negligible. The heat conduction is expressed in terms of Fourier's law and the radiation is taken to be of the diffusion type for an optically thick grey gas model. The thermal conductivity and absorption coefficient are assumed to be proportional to appropriate powers of temperature and density. The assumption of an optically thick grey gas is physically consistent with the neglect of radiation pressure and radiation energy. The viscosity is also assumed to be negligible. The shock is assumed to be isothermal. The motion of the piston or explosion is assumed to follow the exponential law (\cite{rangarao, srivastava, vis1, vis2}), namely,
\begin{equation}\label{4}
r_{p}=B^{*} exp(\lambda t), \lambda>0,
\end{equation}

where $r_{p}$ is the radius of the piston or explosion, $t$ is the time, $\lambda$ is a dimensional constant,  and `$B^{*}$' denotes the radius of the piston at time $t=0$. The law of piston motion (\ref{4}) implies a boundary condition on the gas speed at a piston, that is required in the determination of the problem. It is also assumed that the shock propagation obeys the exponential law
\begin{equation}\label{5}
R=B exp(\lambda t),
\end{equation}

where $R$ is the shock radius, and `$B$' is a dimensional constant which depends on the constant `$B^{*}$' and the non-dimensional position of the piston. As it is often the case in the problems of this type, it is more convenient to solve for the piston motion in terms of the shock motion, rather than vice versa. We shall, therefore, adopt this point of view forthwith, and consider `$B$' a known parameter of the problem, rather `$B^{*}$'.

The effects of variation of the strength of the ambient magnetic field, heat transfer parameters, adiabatic exponent, ambient density variation index, on the shock strength, the distance between the piston and the shock front, and the flow variables such as reduced velocity, reduced density, reduced pressure, reduced azimuthal magnetic field, reduced total heat flux, isothermal speed of sound, reduced adiabatic compressibility are studied. It is shown that the shock strength is independent from the heat transfer parameters and the ambient density variation index. The azimuthal magnetic field, and heat conduction and radiation heat flux have same effects on the distance between the piston and the shock front. The compressibility of the medium is increased in the non-magnetic field. The thermal conductivity $K$ and the absorption coefficient $\alpha_R$ depends on the ambient density variation index $\frac{\alpha}{\lambda}$.

\section{Equations of motion and boundary conditions}
The governing system of equations for the one dimensional unsteady, adiabatic flow of an exponential shock wave in a perfect gas with radiation and heat conduction together with azimuthal magnetic field, can be written as (c.f. \cite{vishu, vis, ghoniem,  gre, chr, sum})
\begin{equation}\label{1.1}
{\frac{\partial \rho}{\partial t}}+u{\frac{\partial \rho}{\partial r}}+\rho {\frac{\partial u}{\partial r}}+{\frac{i u\rho}{r}} = 0,
\end{equation}

\begin{equation}\label{1.2}
\frac{\partial u}{\partial t}+u\frac{\partial u}{\partial r}+\frac{1}{\rho}\biggl[\frac{\partial p}{\partial r}+\mu h \frac{\partial h}{\partial r}+\frac{\mu h^2}{r}\biggr]= 0,
\end{equation}

\begin{equation}\label{1.3}
\frac{\partial h}{\partial t}+u\frac{\partial h}{\partial r}+h\frac{\partial u}{\partial r}+(i-1) \frac{hu}{r} = 0,
\end{equation}

\begin{equation}\label{1.4}
\frac{\partial U_{m}}{\partial t}+u\frac{\partial U_{m}}{\partial r}-\frac{p}{\rho^{2}}\biggl(\frac{\partial \rho}{\partial t}+u\frac{\partial \rho}{\partial r}\biggr)+\frac{1}{\rho r^i}\frac{\partial (r^i F)}{\partial r} = 0,
\end{equation}

where $r$ and $t$ are independent space and time coordinates, $\rho$ is the density, $p$ is the pressure,  $u$ is the fluid velocity, $h$ is the azimuthal magnetic field, $U_m$ is the internal energy per unit mass, $\mu$ is the magnetic permeability and $F$ is the total heat flux, where $i=0,1,2$ for planer, cylindrical and spherically symmetry of the flow field.

The above system of equations should be supplemented with an equation of state. As the behaviour of the gas is taken to be ideal, so that
\begin{equation}\label{1.5}
p=\Gamma \rho T ; \;\;U_m=\frac{p}{(\gamma-1)\rho},
\end{equation}

where $\Gamma$ is gas constant, and $\gamma$ is the adiabatic index.

The total heat flux $F$ which appears in the energy equation (\ref{1.4}) can be written as
\begin{equation}\label{1.6}
F=F_c+F_R,
\end{equation}

where $F_c$ is conduction heat flux and $F_R$ is radiation heat flux. According to Fourier's law of heat conduction, the heat conduction $F_c$ can be expressed as
\begin{equation}\label{1.7}
F_c=-K\frac{\partial T}{\partial r},
\end{equation}

where `$K$' is the coefficient of the thermal conductivity and `$T$' is the absolute temperature of the gas. The radiation heat flux $F_R$ can be obtained from the differential approximation of the radiation transport equation in the diffusion limit by assuming local thermodynamic equilibrium and using the radiative diffusion model for an optically thick grey gas (Pomraning \cite{pomraning}). Therefore, the radiation heat flux $F_R$ can be written as follows
\begin{equation}\label{1.8}
F_R=- \frac{4}{3}\left(\frac{\sigma}{\alpha_R}\right)\frac{\partial T^4}{\partial r},
\end{equation}

where $\sigma$ is the Stefan-Boltzmann constant and $\alpha_R$ is the Rosseland mean absorption coefficient. The thermal conductivity $K$ and the absorption coefficient $\alpha_R$ are assumed to vary with density and temperature. According to power laws, these can be written as follows(c.f. \cite{kk, vis, vis11, ghoniem})
\begin{equation}\label{1.9}
K=K_0\left(\frac{T}{T_0}\right)^{\beta_c}\left(\frac{\rho}{{\rho}_0}\right)^{\delta_c},\;\;\alpha_R=\alpha_{R_0}\left(\frac{T}{T_0}\right)^{\beta_R}\left(\frac{\rho}{{\rho}_0}\right)^{\delta_R},
\end{equation}

where the subscript `$0$' denotes a reference state. For existence of similarity solution, the exponents $\beta_c$, $\delta_c$, $\delta_R$ and $\beta_R$ must satisfy the similarity requirements.

We assume that an exponential shock wave is propagating in the undisturbed ideal gas with variable density in the presence of variable azimuthal magnetic field, conduction and radiation heat flux. The flow variables immediately ahead of the shock front are given by
\begin{align}
u=&u_a=0, \label{1.10}\\
 \rho=&\rho_a=\rho_0 \;\rm{exp}(-\alpha t),\;\;\alpha>0 \label{1.11} \\
 h=&h_a=h_0 \;\rm{exp}(-\delta t),\label{1.12} \\
 F=&F_a=0,\;\;(\text{Laumbach and Probstein \cite{laumbach}}) 
\end{align}
where $\rho_0$, $h_0$, $\alpha$ and $\delta$ are dimensional constants and the subscript `$a$' refers to the condition immediately ahead of the shock front. From equations (\ref{1.2}), (\ref{1.11}) and (\ref{1.12}), we have
\begin{equation}\label{1.13}
p_a=\frac{\alpha {h_0}^2}{2}\;\rm{exp}(-2\delta t)(\frac{\lambda}{\delta}-1),\;\lambda>0 .
\end{equation}

The jump conditions across the shock front are given by the law of conservation of mass, momentum and energy across the isothermal shock (the isothermal shock is formed by the mathematical approximation in which the flux is assumed to be proportional to the temperature gradient. This assumption of isothermal shock excludes the possibility of a temperature jump, see for example \cite{kk, rosenau, vish8, zel}), we have
\begin{align}
&\rho_a V=\rho_n(V-u_n),\label{1.14}\\
&h_a V=h_n(V-u_n),\label{1.15}\\
&p_a+\frac{\mu {h_a}^2}{2}+\rho_a V^2=p_n+\frac{\mu {h_n}^2}{2}+\rho_n(V-u_n)^2,\label{1.16}\\
&e_a+\frac{p_a}{\rho_a}+\frac{V^2}{2}+\frac{\mu {h_a}^2}{\rho_a}+\frac{F_n}{\rho_a V}=e_n+\frac{p_n}{\rho_n}+\frac{(V-u_n)^2}{2}+\frac{\mu {h_n}^2}{\rho_n},\label{1.17}\\
&T_a=T_n,\label{1.18}
\end{align}
where $V=\lambda R$ denotes the velocity of the shock front and the subscript `$n$' denotes the conditions immediately behind the shock front.

The shock conditions (\ref{1.14})-(\ref{1.18}) across the isothermal shock propagating into perfect gas reduces to
\begin{align}
u_n&=(1-\beta)V,\label{1.19}\\
\rho_n&=\frac{\rho_a}{\beta},\label{1.20}\\
h_n&=\frac{h_a}{\beta},\label{1.21}\\
p_n&=\rho_a V^2[1-\beta+\frac{1}{\gamma M^2}+\frac{1}{2 {M_A}^2}(1-\frac{1}{\beta^2})],\label{1.22}\\
F_n&=(1-\beta)\rho_a V^3 [\frac{\beta(\gamma+1)}{2(\gamma-1)}-(\frac{1}{2}+\frac{1}{M^2(\gamma-1)}+\frac{{M_A}^{-2}}{2}\frac{\gamma}{\gamma-1})\nonumber\\
&{}+\frac{{M_A}^{-2}}{2\beta}\frac{(\gamma-2)}{(\gamma-1)}],\label{1.23}
\end{align}
where $M=(\frac{\rho_a V^2}{\gamma p_a})^{1/2}$ is the shock Mach number where the frozen speed of sound is $(\frac{\gamma p_a}{\rho_a})^{1/2}$ and $M_A=(\frac{\rho_a V^2}{\mu h_a})^{1/2}$ is the Alfven-Mach number. It is found that the shock Mach number $M$ and Alfven-Mach number $M_A$ are constants for $\alpha=2(\lambda+\delta)$. The quantity $\beta$ ($0<\beta<1$) is obtained from the relation
\begin{equation}\label{1.24}
\beta^3-\beta^2(1+\frac{1}{\gamma M^2}+\frac{{M_A}^{-2}}{2})+\frac{\beta}{\gamma M^2}+\frac{{M_A}^{-2}}{2}=0.
\end{equation}

\section{Similarity Transformations}
Zel'dovich and Raizer have presented that the gasdynamic equations reveal similarity transformations, that there are feasible distinct flows similar to each other by changing the basic scales of time, length and density. For, self similar motions, the system of fundamental partial differential equations (\ref{1.1})-(\ref{1.4}) reduces to a system of ordinary differential equations in new unknown functions of the similarity variable $\xi$, which is given by
\begin{equation}
\xi=\frac{r}{R},\;R=R(t).\nonumber
\end{equation}

The velocity, density, pressure, azimuthal magnetic field, heat flux and length scales are not all independent of each other. If we choose $R$ and $\rho_a$ as the basic scales, then the quantity $\frac{dR}{dt}=V$ can serve as the velocity scale, $\rho_a {\dot{R}}^2$ as the pressure scale. This does not restrict the generality of the solution because a scale is only defined within a numerical coefficient, which can always be comprised in the new unknown function. Therefore, we present the solution of the partial differential equations (\ref{1.1})-(\ref{1.4}) in terms of products of scale functions and the new unknown functions of the similarity variable $\xi$ as (c.f. \cite{rangarao, na09} )
\begin{align}
&u=V U(\xi),\;\; \rho=\rho_a D(\xi),\;\; p=\rho_a V^2 P(\xi),\label{1.27}\\
&h=\left(\frac{\rho_a}{\mu}\right)^{\frac{1}{2}} V H(\xi),\;\; F=\rho_a V^3 Q(\xi),\nonumber
\end{align}
where $U$, $D$, $P$, $H$ and $Q$ are functions of $\xi$ only. At the shock front $\xi=1$ and at the piston $\xi=\xi_p$.

Using the similarity transformations (\ref{1.27}), the system of partial differential equations (\ref{1.1})-(\ref{1.4}) reduces into 
\begin{align}
&(U-\xi)\frac{dD}{d\xi}+D(\xi)\frac{dU}{d\xi}+i\frac{DU}{\xi}-\frac{\alpha D}{\lambda}=0,\label{1.29}\\
&(U-\xi)\frac{dU}{d\xi}+U(\xi)+\frac{1}{D(\xi)}\left(\frac{dP}{d\xi}+H(\xi)\frac{dH}{d\xi}+\frac{H(\xi)^2}{\xi}\right)=0,\label{1.30}\\
&(U-\xi)\frac{dH}{d\xi}+\left(1-\frac{\alpha}{2\lambda}\right)H(\xi)+H(\xi)\frac{dU}{d\xi}+(i-1)\frac{H(\xi)U(\xi)}{\xi}=0,\label{1.31}
\end{align}
\begin{align}
&(U-\xi)\frac{dP}{d\xi}-\gamma(U-\xi)\frac{P(\xi)}{D(\xi)}\frac{dD}{d\xi}+\left(2+\frac{\alpha(\gamma-1)}{\lambda}\right)P(\xi)+\frac{i(\gamma-1)}{\xi}Q(\xi)\nonumber\\
&{}+(\gamma-1)\frac{dQ}{d\xi}=0.\label{1.32}
\end{align}

By using the equations (\ref{1.7}), (\ref{1.8}), (\ref{1.9}) into the equation (\ref{1.6}), we get the total heat flux as
\begin{equation}\label{1.33}
F=-\frac{K_0}{{T_0}^{\beta_c}}T^{\beta_c} \frac{\partial T}{\partial r}-\frac{16\sigma {T_0}^{\beta_R}}{3{\alpha}_{R_0}}T^{3-\beta_R}\frac{\partial T}{\partial r}.
\end{equation}

By using the equations (\ref{1.5}) and (\ref{1.27}) in the equation (\ref{1.33}), we get the non-dimensional total heat flux $Q$ as
\begin{flalign}\label{1.34}
&Q=[-\frac{K_0 \lambda {\rho_i}^{{\delta_c}-1} (\lambda B)^{2({\beta_c}-1)}}{{T_0}^{\beta_c}{\rho_0}^{\delta_c} \Gamma^{{\beta_c}+1}}\rm{exp}\{-(\alpha({\delta_c}-1)+2\lambda({\beta_c}-1))t\} P^{\beta_c}D^{{\delta_c}-{\beta_c}} \nonumber\\
&{}-\frac{16 \sigma {T_0}^{\beta_R} {\rho_0}^{\delta_R}\lambda {\rho_i}^{-({\delta_R}+1)}(\lambda B)^{2(2-{\beta_R})}}{3{{\alpha}_{R_0}}{\Gamma^{4-{\beta_R}}}}\rm{exp}((\alpha({\delta}+1)+2{\lambda}(2-{\beta_R}))t)\nonumber\\
&{}\times D^{{\beta_R}-{\delta_R}-3} P^{3-{\beta_R}}]\frac{\partial}{\partial \eta}\left(\frac{P}{D}\right).
\end{flalign}

The equation (\ref{1.34}) shows that the similarity solution of the present problem exists only when
\begin{equation}\label{1.35}
\beta_c=1+\frac{\alpha}{2\lambda}(\delta_c-1),\;\;\beta_R=2+\frac{\alpha}{2\lambda}(\delta_R+1).
\end{equation}

These relations show that thermal conductivity $K$ and absorption coefficient $\alpha_R$ depends on the ambient density variation index $\frac{\alpha}{\lambda}$. For the case of constant density, the relation (\ref{1.35}) is similar to the relations (37) of Nath and Sahu \cite{sahu2} and (84) of Bajargaan and Patel \cite{baj}.

Under the above condition (\ref{1.35}), the equation (\ref{1.34}) becomes
 \begin{equation}\label{1.36}
Q=-X\left[\frac{1}{D}\frac{dP}{d\eta}-\frac{P}{D^2}\frac{dD}{d\eta}\right]
\end{equation}
 
where $X={{\Gamma_c}P^{\beta_c}D^{{\delta_c}-{\beta_c}}+{\Gamma_R} D^{{\beta_R}-{\delta_R}-3} P^{3-{\beta_R}}}$, ${\Gamma_c}$ and ${\Gamma_R}$ are the non-dimensional conductive and radiative heat transfer parameters, respectively. The parameters ${\Gamma}_c$ and ${\Gamma}_R$ depend on the thermal conductivity $K$ and the mean free path of radiation $1/\alpha_R$, respectively and also on the shock radius exponent $\lambda$ and the dimensional constant $B$, and they are given by
\begin{flalign}\label{1.37}
{\Gamma}_c=\frac{K_0 \lambda {\rho_i}^{{\delta_c}-1} (\lambda B)^{2({\beta_c}-1)}}{{T_0}^{\beta_c}{\rho_0}^{\delta_c} \Gamma^{{\beta_c}+1}},\;\Gamma_R=\frac{16 \sigma {T_0}^{\beta_R} {\rho_0}^{\delta_R}\lambda {\rho_i}^{-({\delta_R}+1)}(\lambda B)^{2(2-{\beta_R})}}{3{{\alpha}_{R_0}}{\Gamma^{4-{\beta_R}}}}.
\end{flalign}

By solving the set of differential equations (\ref{1.29})-(\ref{1.32}) and (\ref{1.36}) for $\frac{dU}{d\xi}$, $\frac{dH}{d\xi}$, $\frac{dP}{d\xi}$, $\frac{dQ}{d\xi}$, $\frac{dD}{d\xi}$, we have
\begin{align}
&\frac{dU}{d\xi}=-\frac{(U-\xi)}{D}\frac{dD}{d\xi}-\frac{iU}{\xi},\label{1.38}\\
&\frac{dH}{d\xi}=\frac{H}{D}\frac{dD}{d\xi}+\frac{HU}{(U-\xi)\xi}-\frac{H}{(U-\xi)},\label{1.39}\\
&\frac{dP}{d\xi}=\left((U-\xi)^2-\frac{H^2}{D}\right)\frac{dD}{d\xi}+\frac{i(U-\xi)DU}{\xi}-DU-\frac{2 H^2}{\xi},\label{1.40}
\end{align}

\begin{align}
&\frac{dQ}{d\xi}=-\frac{dD}{d\xi}\left[(U-\xi)^2-\frac{H^2}{D}-\frac{\gamma P}{D}\right](\frac{(U-\xi)}{(\gamma-1)})+\frac{DU(U-\xi)}{(\gamma-1)}\nonumber\\
&{}-\frac{iDU(U-\xi)^2}{(\gamma-1)\xi}+\frac{2 H^2(U-\xi)}{(\gamma-1)\xi}-2P+\frac{iQ}{\xi},\label{1.41}\\
&\frac{dD}{d\xi}=\left[\frac{D^2}{P-D^2(U-\xi)^2-H^2 D}\right]\times\left[\frac{i(U-\xi)DU}{\xi}-\frac{DU}{P}-\frac{2 H^2}{\xi}+\frac{Q}{X}\right].\label{1.42}
\end{align}

By using the similarity transformations (\ref{1.27}), the shock conditions (\ref{1.19})-(\ref{1.23}) are transformed into
\begin{align}
&U(1)=1-\beta,\nonumber\\
&D(1)=\frac{1}{\beta},\nonumber\\
&H(1)=\frac{M_A^{-1}}{\beta},\nonumber\\
&P(1)=1-\beta+\frac{1}{\gamma M^2}+\frac{M_A^{-2}}{2}\left(1-\frac{1}{\beta^2}\right),\nonumber\\
&Q(1)=(1-\beta)\left[\frac{\beta(\gamma+1)}{2(\gamma-1)}-\frac{1}{2}-\frac{M^{-2}}{(\gamma-1)}-\frac{\gamma M_A^{-2}}{2(\gamma-1)}+\frac{M_A^{-2}}{2\beta}\times\frac{(\gamma-2)}{(\gamma-1)}\right].\label{1.43}
\end{align}

Along with the shock conditions (\ref{1.43}), the condition which is to be satisfied at the piston surface is that the velocity of the fluid is equal to the velocity of the piston itself. From Eq. (\ref{1.27}), this kinematic condition can be written as
\begin{equation}
U(\xi_p)=\xi_p,\label{1.44}
\end{equation}
where $\xi_p$ is the value of $\xi$ at the piston.

For an isentropic change of state of the perfect gas, we may calculate the isothermal speed of sound in perfect gas as follows
\begin{equation}
a_{iso}=\left(\frac{\partial p}{\partial \rho}\right)^{\frac{1}{2}}_{T}=\left(\frac{\gamma p}{\rho}\right)^{\frac{1}{2}},\label{1.45}
\end{equation}
where the subscript `T' refers to the process of constant temperature.

By using the transformations (\ref{1.27}) in the equation (\ref{1.45}), the expression for reduced isothermal speed of sound is given by
\begin{equation}
\frac{a_{iso}}{\dot{R}}=\left(\frac{P}{D}\right)^{\frac{1}{2}}.\label{1.46}
\end{equation}

The adiabatic compressibility of perfect gas can be calculated as (c.f. Moelwyn-Hughes \cite{moel})
\begin{equation}
C_{adi}=\frac{1}{\rho}\left(\frac{\partial p}{\partial \rho}\right)_s=\frac{1}{\gamma p}.
\end{equation}

The reduced adiabatic compressibility for perfect gas can be written as
\begin{equation}
\frac{C_{adi}}{(C_{adi})_n}=\frac{P(1)}{P(\xi)}.
\end{equation}

The total energy of the flow field between the piston and the shock wave is given by
\begin{equation}\label{1111}
E=2\pi i\int_{r_p}^{R} \rho\left[ U_m+\frac{u^2}{2}+\frac{\mu {h^2}}{2\rho}\right] r^i dr,
\end{equation}
where $r_p$ is the radius of the piston or inner expanding surface.

By using the similarity transformations (\ref{1.27}) and the equation (\ref{1.5}) in the relation (\ref{1111}), we have
\begin{equation}
E=2 \pi {\rho_a} i {\lambda^2} R^{3+i} J,
\end{equation}
where
\begin{equation}
J=\int_{\xi_p}^{1} \left[ \frac{P(\xi)}{(\gamma-1)}+\frac{{U(\xi)}^2 D(\xi)}{2}+\frac{{H(\xi)}^2}{2}\right] \xi^i d\xi,\nonumber
\end{equation}
$\xi_p$ being the value of `$\xi$' at the piston or inner expanding surface.

Thus the total energy of the shock wave is not constant and varies as $R^{3+i}$ where $i=1$ or $2$ for cylindrical or spherical shock wave. The increase in the total energy may be achieved by the pressure exerted on the fluid by the piston.
 
Normalizing the flow variables $u$, $\rho$, $p$, $h$, $F$ and $C_{adi}$ with their respective values at the shock front, we obtain
\begin{eqnarray*}
&&\frac{u}{u_n}=\frac{U(\xi)}{U(1)},\;\;\frac{\rho}{\rho_n}=\frac{D(\xi)}{D(1)},\;\;\frac{p}{p_n}=\frac{P(\xi)}{P(1)},\\
&&\frac{h}{h_n}=\frac{H(\xi)}{H(1)},\;\;\frac{F}{F_n}=\frac{Q(\xi)}{Q(1)},\;\;\frac{C_{adi}}{(C_{adi})_n}=\frac{P(1)}{P(\xi)}.
\end{eqnarray*}

\section{Results and discussion}
For the existence of similarity solution of the present problem, the shock Mach number $M$ and Alfven-Mach number $M_A$ must be constant. Therefore, the solution of the problem exist when the following condition must be satisfied
\begin{equation}\label{567}
\lambda+\delta=\frac{\alpha}{2}>0.
\end{equation}

The distribution of the flow variables between the shock front $(\xi=1)$ and the inner expanding surface or piston $(\xi=\xi_p)$ is obtained by the numerical integration of equations (\ref{1.38})-(\ref{1.42}) with the boundary conditions (\ref{1.43}) by using Runge-Kutta method of the fourth order. For the determination of numerical integration, the values of the constant parameters are taken to be $i=2$ (spherically symmetric flow); $\frac{\alpha}{\lambda}=1.5, 2, 2.5$; $\gamma=\frac{4}{3}, \frac{5}{3}$; $M_A^{-2}=0, 0.01, 0.1$; $M=5$; $\delta_c=1$; $\delta_R=2$; ${\Gamma}_{c}=0.1, 10, 1000$; ${\Gamma}_{R}=0.5, 10, 500$. We have taken three values of ambient density variation index $\frac{\alpha}{\lambda}=1.5, 2, 2.5$ for numerical computations. The three chosen values of $\frac{\alpha}{\lambda}=1.5, 2, 2.5$ corresponds to the increasing, constant and decreasing ambient magnetic field variation index ($\frac{\delta}{\lambda}=-0.25, 0, 0.25$) ahead of the shock front by the equations (\ref{1.12}) and (\ref{567}).For the existence of shocks propagating in regions of variable density, there is a astrophysical evidence. In a stellar explosion, the shock wave is expected to accelerate through the outer stellar layers where the density is decreasing rapidly with height. A similar situation may occur for an explosion in the gaseous atmosphere of a galaxy. We have taken two values of $\gamma$, i.e. $\gamma=\frac{5}{3}$ for fully ionized gas and $\gamma=\frac{4}{3}$ for relativistic gas, which are applicable to interstellar medium. The most general range of values of adiabatic exponent seen in real stars are marked by these two values of $\gamma$. We have taken above values of $M_A^{-2}$ in the present problem because Rosenau and Frankenthal \cite{rosenau} have shown that the effects of magnetic field on the flow-field behind the shock are significant when $M_A^{-2} \geq 0.01$. The value $0$ of $M_A^{-2}$ represents the non-magnetic case. The set of values ${\delta}_{c}=1$, ${\delta}_{R}=2$ is the representative of the case of high-temperature, low density medium (Ghoneim et al. \cite{ghoniem}). Also, the set of values ${\Gamma}_{c}=10$, ${\Gamma}_{R}=0.5$ (taken in Fig. 1(a)-1(g) is the representative of the case in which there is heat transfer by both the conduction and radiative diffusion. 

Figures 1(a)-1(g), 2(a)-2(g) and 3(a)-3(g) show the variation of the reduced flow variables $\frac{u}{u_n}$, $\frac{\rho}{\rho_n}$, $\frac{p}{p_n}$, $\frac{h}{h_n}$, $\frac{F}{F_n}$, $\frac{a_{iso}}{\dot{R}}$, $\frac{C_{adi}}{(C_{adi})_n}$ with the similarity variable $\xi$ at various values of the parameters ${M_A}^{-2}$, $\Gamma_c$, $\Gamma_R$, $\gamma$, $\frac{\alpha}{\lambda}$. As we move towards the shock front from the piston, figures 1(a)-1(g) show that in non-magnetic field, the velocity $\frac{u}{u_n}$ decrease for $\frac{\alpha}{\lambda}=1.5, 2$ but almost constant for $\frac{\alpha}{\lambda}=2.5$; the density $\frac{\rho}{\rho_n}$ and the pressure $\frac{p}{p_n}$ decrease; the total heat flux $\frac{h}{h_n}$, the isothermal speed of sound $\frac{a_{iso}}{\dot{R}}$ and the adiabatic compressibility $\frac{C_{adi}}{(C_{adi})_n}$ increase . Further, in magnetic field, the velocity $\frac{u}{u_n}$ decreases for $\frac{\alpha}{\lambda}=1.5$ but constant for $\frac{\alpha}{\lambda}=2$, ${M_A}^{-2}=0.01$ and increases for the rest cases; the pressure $\frac{p}{p_n}$ and the density $\frac{\rho}{\rho_n}$ decreases rapidly  for ${M_A}^{-2}=0.01$, $\frac{\alpha}{\lambda}=2, 2.5$ and are almost constant for other cases; the adiabatic compressibility $\frac{C_{adi}}{(C_{adi})_n}$ increases rapidly for ${M_A}^{-2}=0.01$, $\frac{\alpha}{\lambda}=2, 2.5$ and has distinct effects for other cases; the magnetic field $\frac{h}{h_n}$, the total heat flux $\frac{F}{F_n}$ increase; the isothermal speed of sound $\frac{a_{iso}}{\dot{R}}$ is almost constant for ${M_A}^{-2}=0.1$, $\frac{\alpha}{\lambda}=2.5$ otherwise increase.

\begin{center}
\bf{Table $1$}
\end{center}
\textbf{Variation of the density ratio $\beta(=\frac{\rho_{a}}{\rho_{n}})$ across the shock front and the position of the piston surface $\xi_{p}$ for different values of $M_A^{-2}$, $\gamma$ and $\frac{\alpha}{\lambda}$ with $M=5$, $\delta_{c}=1$, $\delta_{R}=2$, $\Gamma_{c}=10$ and $\Gamma_{R}=0.5$.}

\begin{center}\small
\begin{tabular}{|l|c|c|c|c|r|}
\hline
 $\gamma$ & $\frac{\alpha}{\lambda}$ & $M_A^{-2}$ & $\beta$ & $1-\beta$ & position of the piston
$\xi_{p}$ \\
\hline
$\frac{4}{3}$ & 1.5 & 0 & 0.03 & 0.97 & 0.98401 \\
  &  & 0.01 & 0.090344 & 0.909656 & 0.94159 \\
    &  & 0.1 & 0.267156  & 0.732844 & 0.811154 \\
   & 2  & 0 & 0.03 & 0.97 &  0.979118 \\
    &  & 0.01  & 0.090344 & 0.909656 &  0.917987 \\
    &  &  0.1 & 0.267156 & 0.732844 &  0.72979 \\
& 2.5 & 0 & 0.03 & 0.97 &  0.966509 \\
    &  & 0.01  & 0.090344 & 0.909656 &  0.87771 \\
    &  &  0.1 & 0.267156 & 0.732844 &  0.536279 \\
$\frac{5}{3}$ & 1.5 & 0 & 0.024 & 0.976 & 0.987802 \\
     & & 0.01 & 0.0866821 & 0.9133179 & 0.942523 \\
     & & 0.1 & 0.263647 & 0.736353 & 0.812244 \\
     & 2 & 0 & 0.024 & 0.976 & 0.979263 \\
     & & 0.01 & 0.0866821 & 0.9133179 & 0.919245 \\
     & & 0.1 & 0.263647 & 0.736353 & 0.735768 \\
& 2.5  & 0 & 0.024 & 0.976 &  0.978603 \\
    &  & 0.01  & 0.0866821 & 0.9133179 &  0.882101 \\
    &  &  0.1 & 0.263647 & 0.736353 &  0.565598 \\
\hline
\end{tabular}
\end{center}

The effects of increase in the value of different flow parameters on the shock propagation are discussed below:

\subsection{Effects of increase in the strength of ambient magnetic field, i.e. effects of increase in the value of \texorpdfstring{${M_A}^{-2}$}:}

The effects of an increase in the value of the strength of ambient magnetic field ${M_A}^{-2}$ in presence of heat conduction and radiation heat flux are manifested as follows:
\begin{enumerate}
\item[(i)] the value of $\beta$ increases i.e. the shock strength ($1-\beta$) decreases (see Table 1);
 \item[(ii)] the position of the piston $\xi_p$ decreases, i.e. the distance between the piston and the shock front ($1-\xi_p$) increases (see Table 1);
\item[(iii)] in change from non-magnetic field $({M_A}^{-2}=0)$ to magnetic field $({M_A}^{-2}>0)$, the velocity $\frac{u}{u_n}$, the density $\frac{\rho}{\rho_n}$ and the pressure $\frac{p}{p_n}$ decrease; the total heat flux $\frac{F}{F_n}$ and the  adiabatic compressibility $\frac{C_{adi}}{(C_{adi})_n}$ increase; and the  isothermal speed of sound $\frac{a_{iso}}{\dot{R}}$ has different effects (see Fig. 1);
\item[(iv)] for the case of magnetic field (${M_A}^{-2}>0$),  the density $\frac{\rho}{\rho_n}$ decreases; the magnetic field $\frac{h}{h_n}$, the total heat flux $\frac{F}{F_n}$ and the  isothermal speed of sound $\frac{a_{iso}}{\dot{R}}$ increase; for $\frac{\alpha}{\lambda}=2, 2.5$, the velocity $\frac{u}{u_n}$ and the pressure $\frac{p}{p_n}$ decrease but for $\frac{\alpha}{\lambda}=1.5$, the pressure $\frac{p}{p_n}$ increase and the velocity $\frac{u}{u_n}$ has negligible effects; further, the adiabatic compressibility $\frac{C_{adi}}{(C_{adi})_n}$ has different variations behind the shock front (see Fig. 1);  This is possible due to the small effects of increasing ambient magnetic field variation index.
\end{enumerate}

It is found that the presence of magnetic field has significant effects on variation of all the flow variables behind the shock front. It is observed that the density decreases with an increase in the strength of ambient magnetic field. Physically it means that gas compressed by shock wave will experience an increase in the strength of ambient magnetic field which is inversely proportional to increase in gas density.

\begin{figure}
\begin{center}
\includegraphics*[height=15cm, width=11cm]{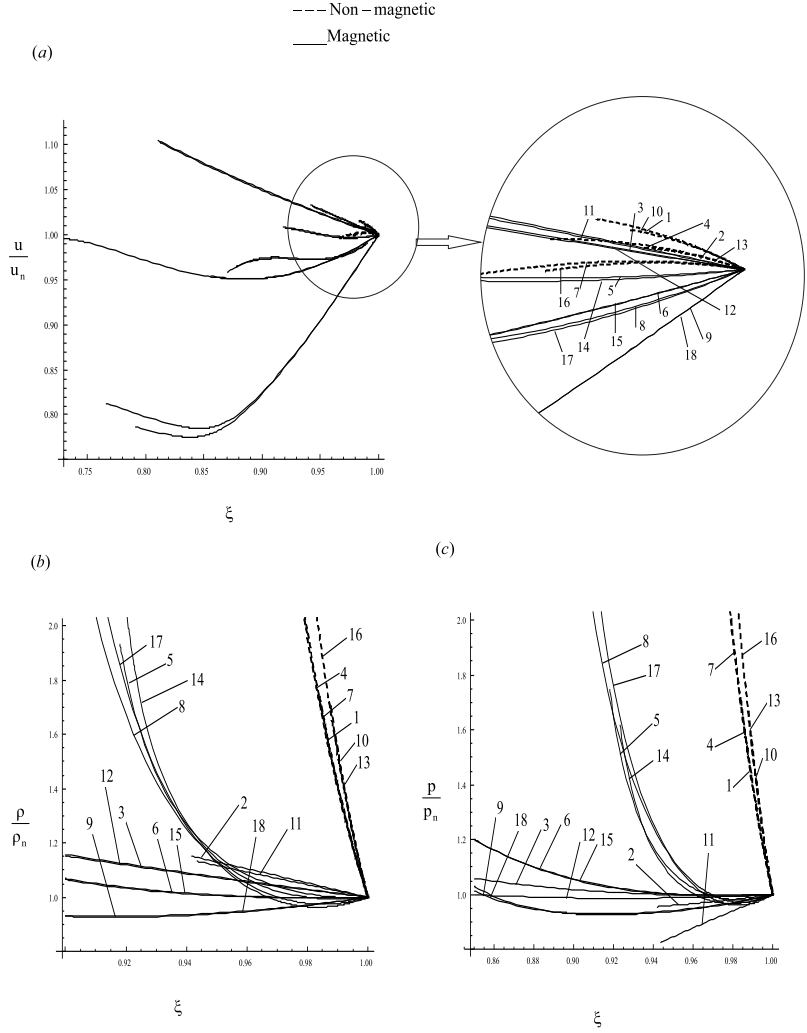}
\end{center}
\caption*{\bf{Fig. 1. Variation of the flow variables (a) reduced velocity (b) reduced density (c) reduced pressure (d) reduced magnetic field, in the region behind the shock front in case of $\Gamma_{c}=10$, $\Gamma_{R}=0.5$, $M=5$, $\delta_c=1$, $\delta_R=2$; 1.$\gamma=\frac{4}{3}$, $\frac{\alpha}{\lambda}=1.5$, ${M_A}^{-2}=0$; 2.$\gamma=\frac{4}{3}$, $\frac{\alpha}{\lambda}=1.5$, ${M_A}^{-2}=0.01$ ; 3.$\gamma=\frac{4}{3}$, $\frac{\alpha}{\lambda}=1.5$, ${M_A}^{-2}=0.1$; 4.$\gamma=\frac{4}{3}$, $\frac{\alpha}{\lambda}=2$, ${M_A}^{-2}=0$; 5.$\gamma=\frac{4}{3}$, $\frac{\alpha}{\lambda}=2$, ${M_A}^{-2}=0.01$; 6.$\gamma=\frac{4}{3}$, $\frac{\alpha}{\lambda}=2$, ${M_A}^{-2}=0.1$; 7.$\gamma=\frac{4}{3}$, $\frac{\alpha}{\lambda}=2.5$, ${M_A}^{-2}=0$; 8.$\gamma=\frac{4}{3}$, $\frac{\alpha}{\lambda}=2.5$, ${M_A}^{-2}=0.01$; 9.$\gamma=\frac{4}{3}$, $\frac{\alpha}{\lambda}=2.5$, ${M_A}^{-2}=0.1$; 10.$\gamma=\frac{5}{3}$, $\frac{\alpha}{\lambda}=1.5$, ${M_A}^{-2}=0$; 11.$\gamma=\frac{5}{3}$, $\frac{\alpha}{\lambda}=1.5$, ${M_A}^{-2}=0.01$; 12.$\gamma=\frac{5}{3}$, $\frac{\alpha}{\lambda}=1.5$, ${M_A}^{-2}=0.1$; 13.$\gamma=\frac{5}{3}$, $\frac{\alpha}{\lambda}=2$, ${M_A}^{-2}=0$; 14.$\gamma=\frac{5}{3}$, $\frac{\alpha}{\lambda}=2$, ${M_A}^{-2}=0.01$; 15.$\gamma=\frac{5}{3}$, $\frac{\alpha}{\lambda}=2$, ${M_A}^{-2}=0.1$; 16.$\gamma=\frac{5}{3}$, $\frac{\alpha}{\lambda}=2.5$, ${M_A}^{-2}=0$; 17.$\gamma=\frac{5}{3}$, $\frac{\alpha}{\lambda}=2.5$, ${M_A}^{-2}=0.01$; 18.$\gamma=\frac{5}{3}$, $\frac{\alpha}{\lambda}=2.5$, ${M_A}^{-2}=0.1$}}
\end{figure}

\begin{figure}
\includegraphics[height=15cm, width=11cm]{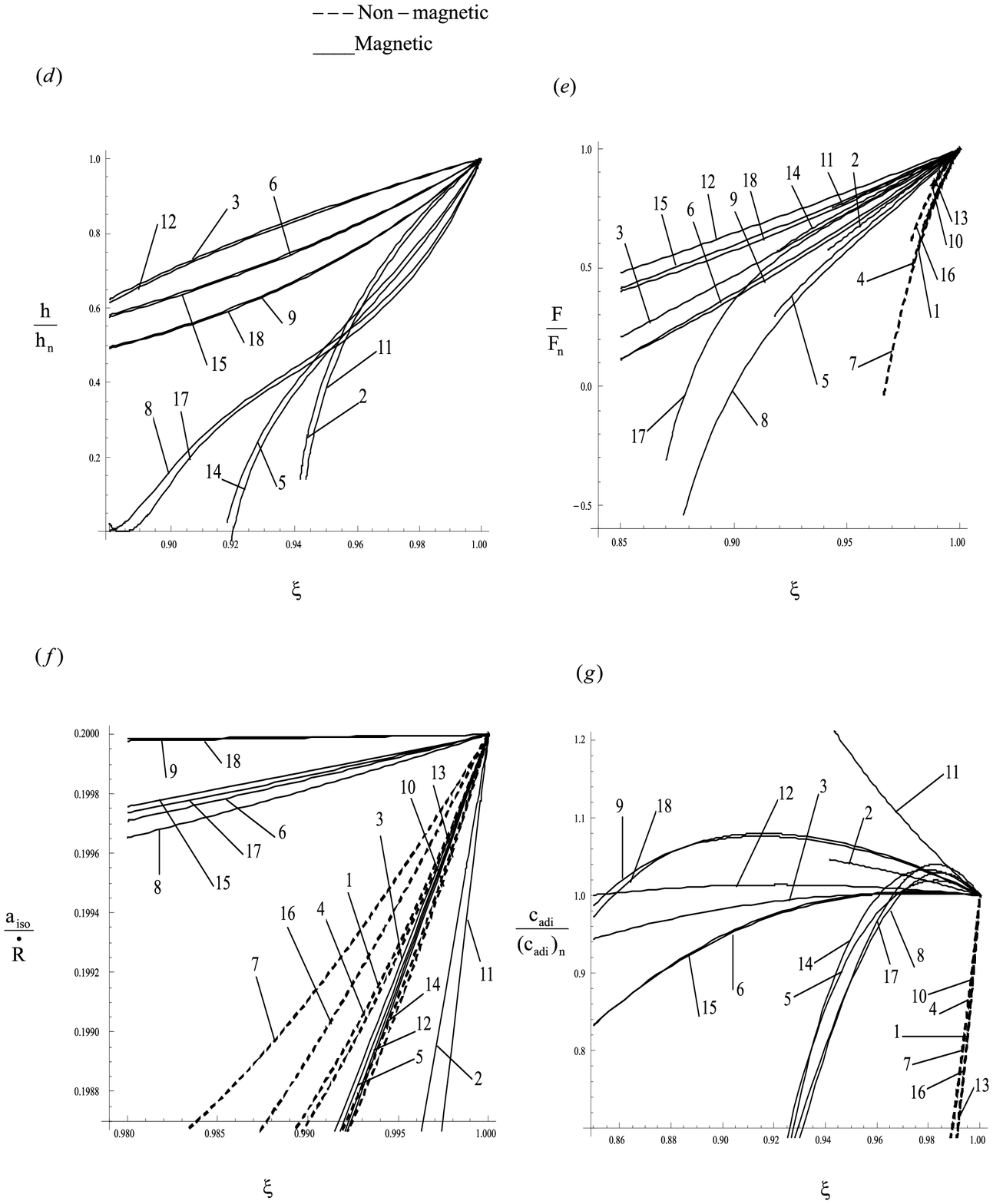}
\caption*{\bf{Fig. 1. Variation of the flow variables (e) reduced total heat flux (f) isothermal speed of sound (g) reduced adiabatic compressibility, in the region behind the shock front in case of $\Gamma_{c}=10$, $\Gamma_{R}=0.5$, $M=5$, $\delta_c=1$, $\delta_R=2$; 1.$\gamma=\frac{4}{3}$, $\frac{\alpha}{\lambda}=1.5$, ${M_A}^{-2}=0$; 2.$\gamma=\frac{4}{3}$, $\frac{\alpha}{\lambda}=1.5$, ${M_A}^{-2}=0.01$ ; 3.$\gamma=\frac{4}{3}$, $\frac{\alpha}{\lambda}=1.5$, ${M_A}^{-2}=0.1$; 4.$\gamma=\frac{4}{3}$, $\frac{\alpha}{\lambda}=2$, ${M_A}^{-2}=0$; 5.$\gamma=\frac{4}{3}$, $\frac{\alpha}{\lambda}=2$, ${M_A}^{-2}=0.01$; 6.$\gamma=\frac{4}{3}$, $\frac{\alpha}{\lambda}=2$, ${M_A}^{-2}=0.1$; 7.$\gamma=\frac{4}{3}$, $\frac{\alpha}{\lambda}=2.5$, ${M_A}^{-2}=0$; 8.$\gamma=\frac{4}{3}$, $\frac{\alpha}{\lambda}=2.5$, ${M_A}^{-2}=0.01$; 9.$\gamma=\frac{4}{3}$, $\frac{\alpha}{\lambda}=2.5$, ${M_A}^{-2}=0.1$; 10.$\gamma=\frac{5}{3}$, $\frac{\alpha}{\lambda}=1.5$, ${M_A}^{-2}=0$; 11.$\gamma=\frac{5}{3}$, $\frac{\alpha}{\lambda}=1.5$, ${M_A}^{-2}=0.01$; 12.$\gamma=\frac{5}{3}$, $\frac{\alpha}{\lambda}=1.5$, ${M_A}^{-2}=0.1$; 13.$\gamma=\frac{5}{3}$, $\frac{\alpha}{\lambda}=2$, ${M_A}^{-2}=0$; 14.$\gamma=\frac{5}{3}$, $\frac{\alpha}{\lambda}=2$, ${M_A}^{-2}=0.01$; 15.$\gamma=\frac{5}{3}$, $\frac{\alpha}{\lambda}=2$, ${M_A}^{-2}=0.1$; 16.$\gamma=\frac{5}{3}$, $\frac{\alpha}{\lambda}=2.5$, ${M_A}^{-2}=0$; 17.$\gamma=\frac{5}{3}$, $\frac{\alpha}{\lambda}=2.5$, ${M_A}^{-2}=0.01$; 18.$\gamma=\frac{5}{3}$, $\frac{\alpha}{\lambda}=2.5$, ${M_A}^{-2}=0.1$ }}
\end{figure}

\newpage
\begin{center}
\bf{Table $2$}
\end{center}
\textbf{Variation of the density ratio $\beta(=\frac{\rho_{a}}{\rho_{n}})$ across the shock front and the position of the piston surface $\xi_{p}$ for different values of $\Gamma_{c}$, $\gamma$ and $\frac{\alpha}{\lambda}$ with $M=5$, $\delta_{c}=1$, $\delta_{R}=2$, $M_A^{-2}=0.01$ and $\Gamma_{R}=10$.}

\begin{center}\small
\begin{tabular}{|l|c|c|c|c|r|}
\hline
 $\gamma$ & $\frac{\alpha}{\lambda}$ & $\Gamma_{c}$ & $\beta$ & $1-\beta$ & position of the piston
$\xi_{p}$ \\
\hline
$\frac{4}{3}$ & 1.5 & 0.1 & 0.090344 & 0.909656  & 0.941053 \\
  &  & 1000 & 0.090344 & 0.909656 & 0.94034 \\
    & 2 & 0.1 & 0.090344  & 0.909656 & 0.915925 \\
   &   & 1000 & 0.090344 & 0.909656 & 0.915042 \\
 & 2.5 & 0.1 & 0.090344  & 0.909656 & 0.842946 \\
   &   & 1000 & 0.090344 & 0.909656 & 0.839723 \\
  $\frac{5}{3}$  & 1.5 & 0.1  & 0.0866821 & 0.9133179 &  942564 \\
    &  &  1000 & 0.0866821 & 0.9133179 &  0.941981 \\
 & 2 & 0.1 & 0.0866821 & 0.9133179 & 0.918722 \\
     & & 1000 & 0.0866821 & 0.9133179 & 0.918168 \\
 & 2.5 & 0.1 & 0.0866821  & 0.9133179 & 0.854114 \\
   &   & 1000 & 0.0866821 & 0.9133179 & 0.854008 \\
  \hline
\end{tabular}
\end{center}

\subsection{Effects of increase in the value of conductive heat transfer parameter \texorpdfstring{$\Gamma_c$} and radiative heat transfer parameter \texorpdfstring{$\Gamma_R$}:}

The effects due to increase in the value of the conductive heat transfer parameter $\Gamma_c$ and the radiative heat transfer parameter $\Gamma_R$ on the shock propagation can be summarized as follows:
\begin{enumerate}
\item[(i)] the shock strength $(1-\beta)$ is independent from the conductive heat transfer parameter $\Gamma_c$ and radiative heat transfer parameter $\Gamma_R$ (see Table 2,3);
\item[(ii)] the distance between the piston and the shock front ($1-\xi_p$) increases by increasing the value of $\Gamma_c$ and $\Gamma_R$ (see Table 2,3);
\item[(iii)] the flow variables have different effects for different values of  $\frac{\alpha}{\lambda}$ due to increase in the value of  $\Gamma_c$ and $\Gamma_R$. By increasing the value of  $\Gamma_c$ , the  velocity $\frac{u}{u_n}$ and the magnetic field $\frac{h}{h_n}$ have negligible effects; for $\frac{\alpha}{\lambda}=1.5$, the density $\frac{\rho}{\rho_n}$, the total heat flux $\frac{F}{F_n}$ and the adiabatic compressibility $\frac{C_{adi}}{(C_{adi})_n}$ decrease but they are almost constant for other values of $\frac{\alpha}{\lambda}$; the pressure $\frac{p}{p_n}$ increases for $\frac{\alpha}{\lambda}=1.5$ otherwise have negligible effects; the isothermal speed of sound increases for $\frac{\alpha}{\lambda}=1.5, 2$ otherwise has negligible effects for $\frac{\alpha}{\lambda}=2.5$ (see Fig 2);
\item[(iv)] by increasing the value of $\Gamma_R$, the flow variables have different behaviour for different values of  $\frac{\alpha}{\lambda}$. The magnetic field $\frac{h}{h_n}$ has negligible effects and the isothermal speed of sound increases; for $\frac{\alpha}{\lambda}=1.5, 2$, the velocity $\frac{u}{u_n}$, the density $\frac{\rho}{\rho_n}$, the total heat flux $\frac{F}{F_n}$ and the adiabatic compressibility $\frac{C_{adi}}{(C_{adi})_n}$ decrease but they have negligible effects for $\frac{\alpha}{\lambda}=2.5$; the pressure increases for $\frac{\alpha}{\lambda}=1.5, 2$ but almost constant for $\frac{\alpha}{\lambda}=2.5$ (see Fig. 3).
\end{enumerate}

The conductive heat transfer parameter $\Gamma_c$ and the radiative heat transfer parameter $\Gamma_R$ have decaying effects on the velocity, the total heat flux, the density, the magnetic field and adiabatic compressibility and these effects are more significant for increasing ambient magnetic field variation index and negligible for decreasing ambient magnetic field variation index. These decaying effects are due to the increase in distance of the piston from the shock front. From equation ($\ref{1.37}$), the increase in value of $\Gamma_c$ and $\Gamma_R$ increase the value of $\lambda$ and hence ($R-r_p$) increases from equation (\ref{4}) and (\ref{5}).

\begin{center}
\bf{Table $3$}
\end{center}
\textbf{Variation of the density ratio $\beta(=\frac{\rho_{a}}{\rho_{n}})$ across the shock front and the position of the piston surface $\xi_{p}$ for different values of $\Gamma_{R}$, $\gamma$ and $\frac{\alpha}{\lambda}$ with $M=5$, $\delta_{c}=1$, $\delta_{R}=2$, $M_A^{-2}=0.01$ and $\Gamma_{c}=10$.}

\begin{center}\small
\begin{tabular}{|l|c|c|c|c|r|}
\hline
 $\gamma$ & $\frac{\alpha}{\lambda}$ & $\Gamma_{R}$ & $\beta$ & $1-\beta$ & position of the piston
$\xi_{p}$ \\
\hline
$\frac{4}{3}$ & 1.5 & 0.5 & 0.090344 & 0.909656  & 0.940594 \\
  &  & 500 & 0.090344 & 0.909656 & 0.940351 \\
    & 2 & 0.5 & 0.090344  & 0.909656 & 0.917933 \\
   &   & 500 & 0.090344 & 0.909656 & 0.915595 \\
& 2.5 & 0.5 & 0.090344  & 0.909656 & 0.84771 \\
   &   & 500 & 0.090344 & 0.909656 & 0.839681 \\
  $\frac{5}{3}$  & 1.5 & 0.5  & 0.0866821 & 0.9133179 &  0.942523 \\
    &  &  500 & 0.0866821 & 0.9133179 &  0.940978 \\
 & 2 & 0.5 & 0.0866821 & 0.9133179 & 0.919245 \\
     & & 500 & 0.0866821 & 0.9133179 & 0.918796 \\
 & 2.5 & 0.5 & 0.0866821 & 0.9133179 & 0.862101 \\
     & & 500 & 0.0866821 & 0.9133179 & 0.853966 \\
  \hline
\end{tabular}
\end{center}

\begin{figure}
\begin{center}
\includegraphics[scale=0.60]{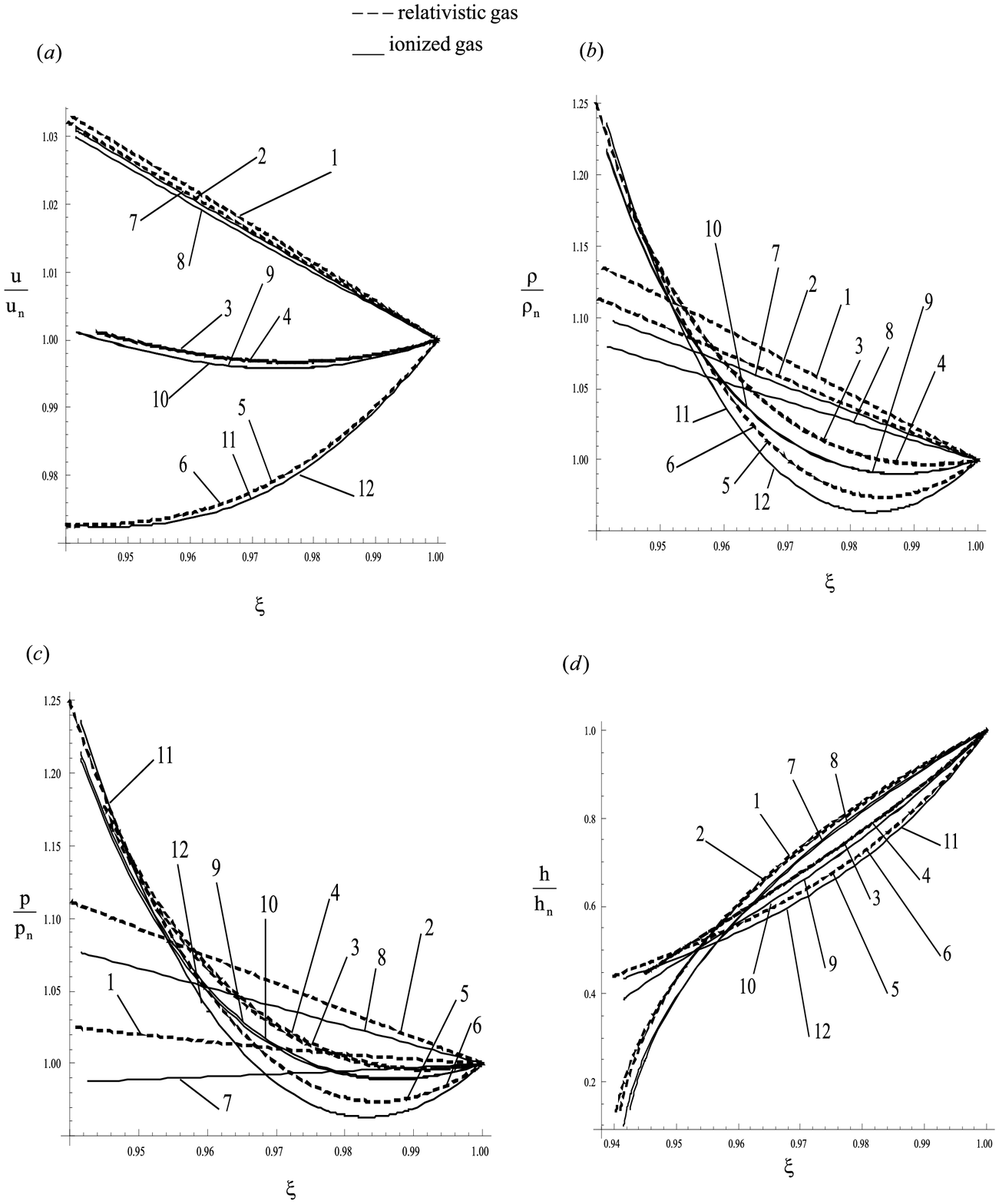}
\end{center}
\caption*{\bf{Fig. 2 Variation of the flow variables (a) reduced velocity (b) reduced density (c) reduced pressure (d) reduced magnetic field, in the region behind the shock front in case of ${M_A}^{-2}=0.01$, $\Gamma_{R}=10$, $M=5$, $\delta_c=1$, $\delta_R=2$; 1.$\gamma=\frac{4}{3}$, $\frac{\alpha}{\lambda}=1.5$, $\Gamma_{c}=0.1$; 2.$\gamma=\frac{4}{3}$, $\frac{\alpha}{\lambda}=1.5$, $\Gamma_{c}=1000$ ; 3.$\gamma=\frac{4}{3}$, $\frac{\alpha}{\lambda}=2$, $\Gamma_{c}=0.1$; 4.$\gamma=\frac{4}{3}$, $\frac{\alpha}{\lambda}=2$, $\Gamma_{c}=1000$; 5.$\gamma=\frac{4}{3}$, $\frac{\alpha}{\lambda}=2.5$, $\Gamma_{c}=0.1$; 6.$\gamma=\frac{4}{3}$, $\frac{\alpha}{\lambda}=2.5$, $\Gamma_{c}=1000$; 7.$\gamma=\frac{5}{3}$, $\frac{\alpha}{\lambda}=1.5$, $\Gamma_{c}=0.1$; 8.$\gamma=\frac{5}{3}$, $\frac{\alpha}{\lambda}=1.5$, $\Gamma_{c}=1000$; 9.$\gamma=\frac{5}{3}$, $\frac{\alpha}{\lambda}=2$, $\Gamma_{c}=0.1$; 10.$\gamma=\frac{5}{3}$, $\frac{\alpha}{\lambda}=2$, $\Gamma_{c}=1000$; 11.$\gamma=\frac{5}{3}$, $\frac{\alpha}{\lambda}=2.5$, $\Gamma_{c}=0.1$; 12.$\gamma=\frac{5}{3}$, $\frac{\alpha}{\lambda}=2.5$, $\Gamma_{c}=1000$}}
\end{figure}

\begin{figure}
\begin{center}
\includegraphics[scale=0.60]{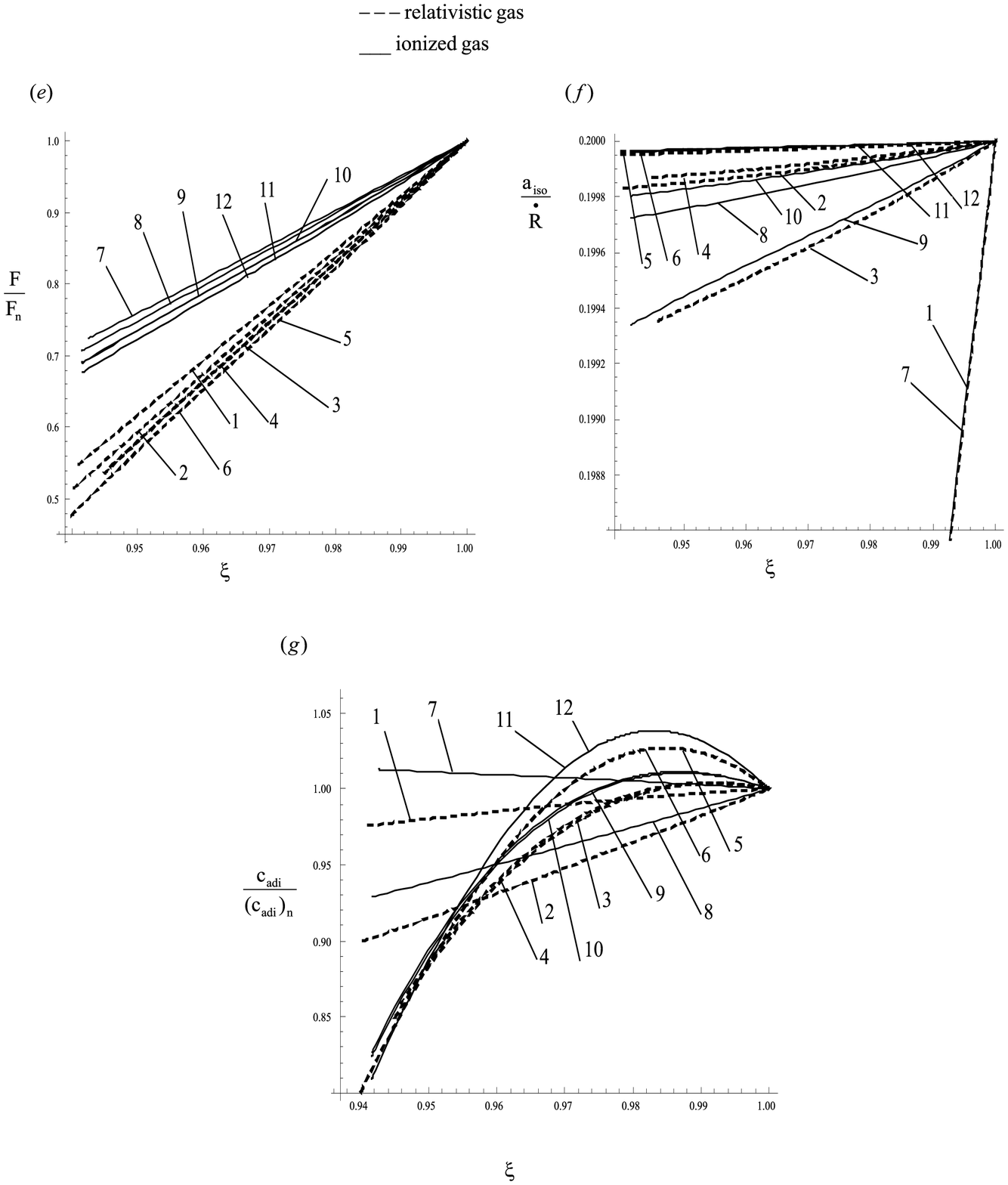}
\end{center}
\caption*{\bf{Fig. 2 Variation of the flow variables (e) reduced total heat flux (f) isothermal speed of sound (g) reduced adiabatic compressibility, in the region behind the shock front in case of ${M_A}^{-2}=0.01$, $\Gamma_{R}=10$, $M=5$, $\delta_c=1$, $\delta_R=2$; 1.$\gamma=\frac{4}{3}$, $\frac{\alpha}{\lambda}=1.5$, $\Gamma_{c}=0.1$; 2.$\gamma=\frac{4}{3}$, $\frac{\alpha}{\lambda}=1.5$, $\Gamma_{c}=1000$ ; 3.$\gamma=\frac{4}{3}$, $\frac{\alpha}{\lambda}=2$, $\Gamma_{c}=0.1$; 4.$\gamma=\frac{4}{3}$, $\frac{\alpha}{\lambda}=2$, $\Gamma_{c}=1000$; 5.$\gamma=\frac{4}{3}$, $\frac{\alpha}{\lambda}=2.5$, $\Gamma_{c}=0.1$; 6.$\gamma=\frac{4}{3}$, $\frac{\alpha}{\lambda}=2.5$, $\Gamma_{c}=1000$; 7.$\gamma=\frac{5}{3}$, $\frac{\alpha}{\lambda}=1.5$, $\Gamma_{c}=0.1$; 8.$\gamma=\frac{5}{3}$, $\frac{\alpha}{\lambda}=1.5$, $\Gamma_{c}=1000$; 9.$\gamma=\frac{5}{3}$, $\frac{\alpha}{\lambda}=2$, $\Gamma_{c}=0.1$; 10.$\gamma=\frac{5}{3}$, $\frac{\alpha}{\lambda}=2$, $\Gamma_{c}=1000$; 11.$\gamma=\frac{5}{3}$, $\frac{\alpha}{\lambda}=2.5$, $\Gamma_{c}=0.1$; 12.$\gamma=\frac{5}{3}$, $\frac{\alpha}{\lambda}=2.5$, $\Gamma_{c}=1000$}}
\end{figure}

\newpage
\subsection{Effects of increase in value of adiabatic exponent \texorpdfstring{$\gamma$}:}

The effects of an increase in the value of adiabatic exponent $\gamma$ are manifested as follows:
\begin{enumerate}
\item[(i)] the shock strength ($1-\beta$) increases (see Table 1, 2, 3);
\item[(ii)] the distance between the piston and the shock front ($1-\xi_p$) decreases (see Table 1, 2, 3);
\item[(iii)] in non-magnetic field, the velocity $\frac{u}{u_n}$ and the adiabatic compressibility $\frac{C_{adi}}{(C_{adi})_n}$ have negligible effects; the density $\frac{\rho}{\rho_n}$, the pressure $\frac{p}{p_n}$ and the total heat flux $\frac{F}{F_n}$ increase in small amount; the isothermal speed of sound $\frac{a_{iso}}{\dot{R}}$ has negligible effects but decreases for $\frac{\alpha}{\lambda}=2.5$ (see Fig. 1);
\item[(iv)] in magnetic field (see Fig. 2), the velocity $\frac{u}{u_n}$, the density $\frac{\rho}{\rho_n}$, the pressure $\frac{p}{p_n}$ and the magnetic field decrease in small amount; the total heat flux $\frac{F}{F_n}$ and the adiabatic compressibility $\frac{C_{adi}}{(C_{adi})_n}$ increase; further, the isothermal speed of sound $\frac{a_{iso}}{\dot{R}}$ has distinct effects.
\end{enumerate}

It is found that the effects of increase in the value of adiabatic exponent $\gamma$ are more impressive in presence of magnetic field than in non-magnetic field. Also, the strength of ambient magnetic field and  the adiabatic exponent have opposite effects on the distance between the piston and the shock front, and on the shock strength. From equation (\ref{1.24}), the increase in value of $\gamma$ decreases the value of $\beta$, therefore the shock strength $(1-\beta)$ increases and we get above effects.
 
\begin{figure}
\begin{center}
\includegraphics[scale=0.60]{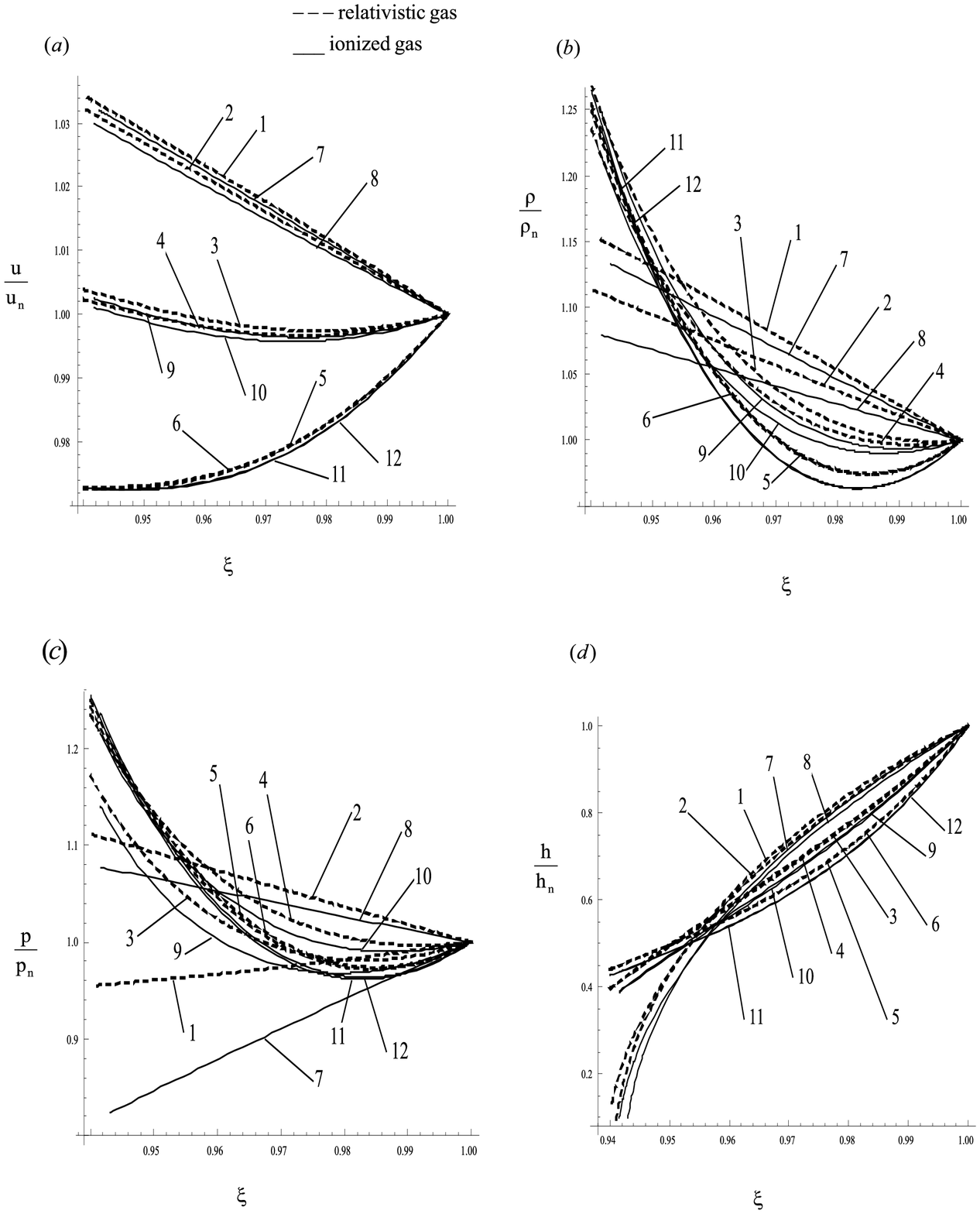}
\end{center}
\caption*{\bf{Fig. 3 Variation of the flow variables (a) reduced velocity (b) reduced density (c) reduced pressure (d) reduced magnetic field, in the region behind the shock front in case of ${M_A}^{-2}=0.01$, $\Gamma_{c}=10$, $M=5$, $\delta_c=1$, $\delta_R=2$; 1.$\gamma=\frac{4}{3}$, $\frac{\alpha}{\lambda}=1.5$, $\Gamma_{R}=0.5$; 2.$\gamma=\frac{4}{3}$, $\frac{\alpha}{\lambda}=1.5$, $\Gamma_{R}=500$ ; 3.$\gamma=\frac{4}{3}$, $\frac{\alpha}{\lambda}=2$, $\Gamma_{R}=0.5$; 4.$\gamma=\frac{4}{3}$, $\frac{\alpha}{\lambda}=2$, $\Gamma_{R}=500$; 5.$\gamma=\frac{4}{3}$, $\frac{\alpha}{\lambda}=2.5$, $\Gamma_{R}=0.5$; 6.$\gamma=\frac{4}{3}$, $\frac{\alpha}{\lambda}=2.5$, $\Gamma_{R}=500$; 7.$\gamma=\frac{5}{3}$, $\frac{\alpha}{\lambda}=1.5$, $\Gamma_{R}=0.5$; 8.$\gamma=\frac{5}{3}$, $\frac{\alpha}{\lambda}=1.5$, $\Gamma_{R}=500$; 9.$\gamma=\frac{5}{3}$, $\frac{\alpha}{\lambda}=2$, $\Gamma_{R}=0.5$; 10.$\gamma=\frac{5}{3}$, $\frac{\alpha}{\lambda}=2$, $\Gamma_{R}=500$; 11.$\gamma=\frac{5}{3}$, $\frac{\alpha}{\lambda}=2.5$, $\Gamma_{R}=0.5$; 12.$\gamma=\frac{5}{3}$, $\frac{\alpha}{\lambda}=2.5$, $\Gamma_{R}=500$}}
\end{figure}

\begin{figure}
\begin{center}
\includegraphics[scale=0.60]{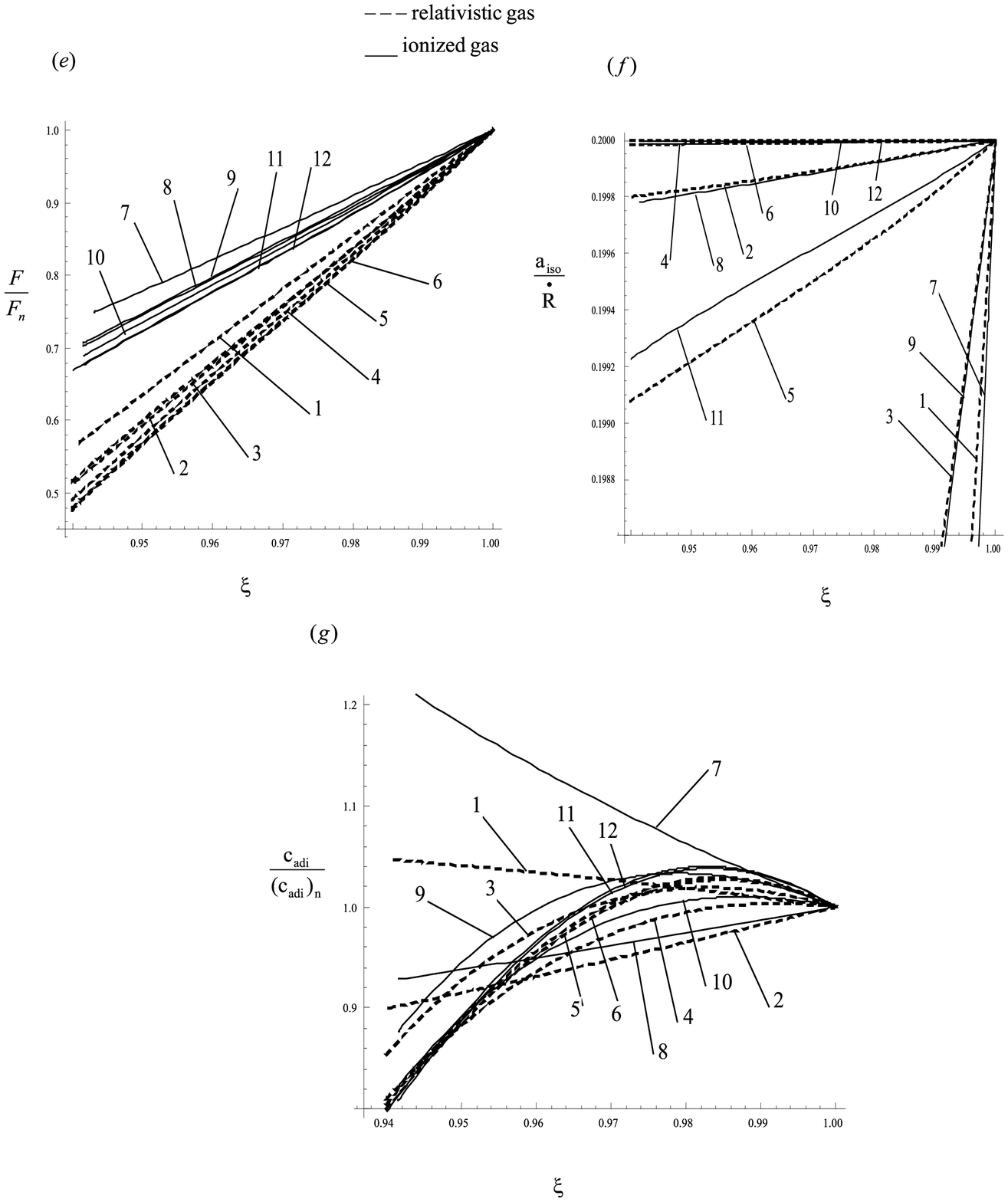}
\end{center}
\caption*{\bf{Fig. 3 Variation of the flow variables (e) reduced total heat flux (f) isothermal speed of sound (g) reduced adiabatic compressibility, in the region behind the shock front in case of ${M_A}^{-2}=0.01$, $\Gamma_{c}=10$, $M=5$, $\delta_c=1$, $\delta_R=2$; 1.$\gamma=\frac{4}{3}$, $\frac{\alpha}{\lambda}=1.5$, $\Gamma_{R}=0.5$; 2.$\gamma=\frac{4}{3}$, $\frac{\alpha}{\lambda}=1.5$, $\Gamma_{R}=500$ ; 3.$\gamma=\frac{4}{3}$, $\frac{\alpha}{\lambda}=2$, $\Gamma_{R}=0.5$; 4.$\gamma=\frac{4}{3}$, $\frac{\alpha}{\lambda}=2$, $\Gamma_{R}=500$; 5.$\gamma=\frac{4}{3}$, $\frac{\alpha}{\lambda}=2.5$, $\Gamma_{R}=0.5$; 6.$\gamma=\frac{4}{3}$, $\frac{\alpha}{\lambda}=2.5$, $\Gamma_{R}=500$; 7.$\gamma=\frac{5}{3}$, $\frac{\alpha}{\lambda}=1.5$, $\Gamma_{R}=0.5$; 8.$\gamma=\frac{5}{3}$, $\frac{\alpha}{\lambda}=1.5$, $\Gamma_{R}=500$; 9.$\gamma=\frac{5}{3}$, $\frac{\alpha}{\lambda}=2$, $\Gamma_{R}=0.5$; 10.$\gamma=\frac{5}{3}$, $\frac{\alpha}{\lambda}=2$, $\Gamma_{R}=500$; 11.$\gamma=\frac{5}{3}$, $\frac{\alpha}{\lambda}=2.5$, $\Gamma_{R}=0.5$; 12.$\gamma=\frac{5}{3}$, $\frac{\alpha}{\lambda}=2.5$, $\Gamma_{R}=500$
}}
\end{figure}

\subsection{Effects of increase in value of ambient density variation index \texorpdfstring{$\frac{\alpha}{\lambda}$}:}

The effects of an increase in the value of ambient density variation index $\frac{\alpha}{\lambda}$ are given as follows:
\begin{enumerate}
\item[(i)] the shock strength $(1-\beta)$ is independent from ambient density variation index $\frac{\alpha}{\lambda}$ (see Table 1, 2, 3);
\item[(ii)]  the distance between the piston and the shock front ($1-\xi_p$) increases (see Table 1, 2, 3);
\item[(iii)] in non-magnetic field, the velocity $\frac{u}{u_n}$ decreases; the density $\frac{\rho}{\rho_n}$, the pressure $\frac{p}{p_n}$, the total heat flux $\frac{F}{F_n}$ and the adiabatic compressibility $\frac{C_{adi}}{(C_{adi})_n}$ have negligible effects; the isothermal speed of sound $\frac{a_{iso}}{\dot{R}}$ has negligible effects for moving from $\frac{\alpha}{\lambda}=1.5$ to $\frac{\alpha}{\lambda}=2$ but increases from $\frac{\alpha}{\lambda}=2$ to $\frac{\alpha}{\lambda}=2.5$ (see fig. 1);
\item[(iv)] in presence of magnetic field (see fig. 2), the velocity $\frac{u}{u_n}$ decreases; the density $\frac{\rho}{\rho_n}$ decreases near the shock front; the magnetic field $\frac{h}{h_n}$ decreases away the piston; the total heat flux $\frac{F}{F_n}$ decreases in small amount; the isothermal speed of sound $\frac{a_{iso}}{\dot{R}}$ increases; the pressure $\frac{p}{p_n}$ and the adiabatic compressibility $\frac{C_{adi}}{(C_{adi})_n}$ increase with $\frac{\alpha}{\lambda}$ upto a certain distance from the shock front and after that behave differently.
\end{enumerate}

It is found that the effects of increase in the value of ambient density variation index are more significant in magnetic field. The heat transfer parameters and the ambient density variation index have same effects on the distance between the piston and the shock front, and on the shock strength.
   
 \section{Conclusions}
The present work investigates the one-dimensional unsteady adiabatic self similar flow behind an exponential shock wave propagating in a perfect gas with azimuthal magnetic field, heat conduction and radiation heat flux. The density and the magnetic field ahead of the shock front are assumed to vary as an exponential law. The effects of variation of the heat transfer parameters, strength of ambient magnetic field, adiabatic exponent and ambient density variation index are investigated on the flow-field behind the shock front. The shock wave in perfect gas with heat conduction and radiation heat flux, variable ambient density and magnetic field can be important for description of shocks in supernova explosions, in the study of a flare produced shock in solar wind, nuclear explosions and central part of star burst galaxies etc. On the basis of this work, one may draw the following conclusions:
\begin{enumerate}
\item[(i)] The findings of the present work provide a clear picture which show that the presence of the heat conduction and radiation heat flux, the ambient variable density and the ambient variable magnetic field brings a profound change in the behaviour of the flow-field behind the shock wave.
\item[(ii)] The similarity solution of the present problem exists only when the sum of shock radius exponent and ambient magnetic field exponent is equal to the half of the ambient density exponent.
\item[(iii)] The total energy of the flow field behind the shock wave is not constant but varies as power of shock radius i.e. $R^{3+i}$ where $i=1$ for cylindrical and $i=2$ for spherical shock wave. 
\item[(iv)] The shock strength decreases by increasing the strength of ambient magnetic field and increases by increasing the value of adiabatic exponent. But, it is independent from the radiative heat transfer parameter, conductive heat transfer parameter and the ambient density variation index.
\item[(v)] The distance between the piston and the shock front increases by increasing the value of ambient magnetic field, the conductive heat transfer parameter and radiative heat transfer parameter, ambient density variation index and it decreases by increasing the value of adiabatic exponent.
\item[(vi)] The flow variables have distinct effects in non-magnetic field and magnetic field by increasing the values of the strength of ambient magnetic field, the conductive heat transfer parameter, the radiative heat transfer parameter, the adiabatic exponent and the ambient density variation index. These effects are more significant in magnetic field.
\item[(vii)] The flow variables have distinct effects for increasing, constant and decreasing ambient magnetic field variation index by increasing the values of the strength of ambient magnetic field, the conductive heat transfer parameter, the radiative heat transfer parameter, the adiabatic exponent and the ambient density variation index. These effects are negligible for increasing ambient magnetic field variation index by increasing the strength of ambient magnetic field. By increasing the values of the conductive heat transfer parameter and the radiative heat transfer parameter, the flow variables have significant effects for increasing ambient magnetic field variation index and negligible effects for decreasing ambient magnetic field variation index.
\end{enumerate}

\section*{Acknowledgement}The research of the first author (Ruchi Bajargaan) is supported by CSIR, New Delhi, India vide letter no. 09/045(1264)/2012-EMR-I. The corresponding author (Arvind Patel) thanks to the University of Delhi, Delhi, India for the R\&D grant vide letter no. RC/2015/9677 dated Oct. 15, 2015. The research of the third author (Manoj Singh) is supported by UGC, New Delhi, India vide letter no. Sch. No./JRF/AA/139/F-297/2012-13 dated January 22, 2013.


\begin{thebibliography}{99}\normalsize\baselineskip 12pt

\bibitem{gv} G. Nath, J.P. Vishwakarma, Similarity solution for the flow behind a shock wave in a non-ideal gas with heat conduction and radiation heat-flux in magnetogasdynamics, commun. Nonlinear Sci. Numer. Simul. 19(5)(2014) 1347-1365. 

\bibitem{marshak} R.E. Marshak, Effect of radiation on shock wave behaviour, Phys. Fluids 1 (1958) 24-29.

\bibitem[Elliot(1960)]{Elliot} L.A. Elliott, Similarity methods in radiation hydrodynamics, Proc. Roy. Soc. London. Ser. A 258 (1960) 287-301.

\bibitem[Wang(1964)]{wang} K.C. Wang, The piston problem with thermal radiation, J. Fluid Mech. 20 (1964) 447-455. J.

\bibitem[]{ru} R. Bajargaan, A. Patel, Similarity solution for a cylindrical shock wave in a self-gravitating, rotating axisymmetric dusty gas with heat conduction and radiation heat flux, J. Appl. Fluid Mech. 10(1) (2017) 329-341.

\bibitem[Nath\; and \;Sahu(2016)]{sahu2} G. Nath, P.K. Sahu, Flow behind an exponential shock wave in a rotational axisymmetric non-ideal gas with conduction and radiation heat flux, Int. J. Appl. Comp. Math. 3(4) (2017) 2785-2801.

\bibitem[Bajargaan\; and \;Patel(2018)]{baj} R. Bajargaan, A. patel, Self similar flow behind an exponential shock wave in a self-gravitating, rotating, axisymmetric dusty gas with heat conduction and radiation heat flux, Indian J. phys. (2018) (in press).

\bibitem[Vishwakarma \;and \; Nath(2011)]{vishu} J.P. Vishwakarma, G. Nath, Cylindrical shock wave generated by a piston moving in a non-uniform self-gravitating
rotational axisymmetric gas in the presence of conduction and radiation heat-flux, Adv. Eng. Res. 2 (2011) 537.

\bibitem[Vishwakarma\;et\;al.(2008)]{kk} J.P. Vishwakarma, G. Nath and K.K. Singh, Propagation of Shock Waves in a Dusty Gas with Heat Conduction, Radiation Heat Flux and Exponentially Varying Density, Phys. Scripta $\bf{78}$ (2008) 11.

\bibitem[Vishwakarma \;and \; Nath(2008)]{vis}J.P. Vishwakarma, G. Nath, Propagation of shock waves in an exponential medium with heat conduction and radiation heat flux, {MMC\_B} 77 (2008) 67-84.

\bibitem[Vishwakarma \;and \; Nath(2010)]{vis11}J.P. Vishwakarma, G. Nath, Propagation of a cylindrical shock wave in a rotating dusty gas with heat conduction and radiation heat flux, Phys. Scripta 81(4) (2010) 045401.


\bibitem[Hartmann(1998)]{hart} L. Hartmann, Accretion processes in star formation, Cambridge University press, Cambridge, 1998.

\bibitem[Balick\; and\;Frank(2002)]{bali} B. Balick, A. Frank, Shapes and Shaping of planetary nebulae, Annu. Rev. Astron. Astrophys. 40 (2002) 439-486.

\bibitem[Sedov(1959)]{Sedov} L.I. Sedov, Similarity and Dimensional Methods in Mechanics, Academic Press, New York, 1959.

\bibitem[Ranga\; Rao\; and\; Ramana(1976)]{rangarao} M.P. Ranga Rao, B.V. Ramana, Unsteady flow of a gas behind an exponential shock, J. Math. Phys. Sci. 10 (1976) 465-476.

\bibitem[Singh\; and \; Srivastava(1988)]{srivastava} V.K. Singh, G.K. Srivastava, Propagation of exponential shock waves in magnetogasdynamics, Astrophys. Space Sci. 155 (1988) 215-224.

\bibitem[Vishwakarma\; and \; Nath(2006)]{vis1} J.P. Vishwakarma, G. Nath, Similarity solutions for unsteady flow behind an exponential shock in a dusty gas, Phys. Scripta 74 (2006) 493-498.

\bibitem[Vishwakarma\; and \; Nath(2007)]{vis2} J.P. Vishwakarma, G. Nath G, Similarity solutions for the flow behind an exponential shock in a non-ideal gas, Meccanica 42 (2007) 331-339.

\bibitem[Singh,\;Husain\;and\;Singh(2011)]{sin} L.P. Singh, A. Husain, M. Singh, A self-similar solution of exponential shock waves in non-ideal magnetogasdynamics, Meccanica 46 (2011) 437-445.

\bibitem[Nath(2015)] {na1} G. Nath, Similarity solutions for unsteady flow behind an exponential shock in an axisymmetric rotating non-ideal gas, Meccanica 50 (2015) 1701-1715.

\bibitem[Nath\; and \;Sahu(2016)]{sahu1} G. Nath, P.K. Sahu, Flow behind an exponential shock wave in a rotational axisymmetric perfect gas with magnetic field and variable density, Springerplus 5 (2016) 1509. 


\bibitem[Nath\; and \;Singh(2017)]{sumeeta} G. Nath, S. Singh, Flow behind magnetogasdynamic exponential shock wave in self-gravitating gas, Int. J. Non Linear Mech. 88 (2017) 102-108.



\bibitem[Ghoneim et. al.(1982)]{ghoniem} A.F. Ghoniem, M.M. Kamel, S.A. Berger, A.K. Oppenheim, Effects of internal heat transfer on the structure of self-similar blast waves, J. Fluid Mech. 117 (1982) 473-491.



\bibitem[Gretler \;and \;Wehle(1993)]{gre} W. Gretler, P. Wehle, Propagation of blast waves with exponential heat release and internal heat conduction and thermal radiation, Shock Waves 3 (1993) 95-104.


\bibitem {chr} A.H. Christer, J.B. Helliwell, Cylindrical shock and detonation waves in magnetogasdynamics, J. Fluid Mech. 39 (1969) 705-725.

\bibitem[Summers(1975)]{sum} D. Summers, An idealized model of a magnetohydrodynamic spherical blast wave applied to a flare produced shock in the solar wind, Astron. Astophys. 45 (1975) 151-158.



\bibitem[Pomraning(1973)]{pomraning} G.C. Pomraning, The Equations of Radiation Hydrodynamics, Int. Series Monographs Natur. Phil. 54 (1973).



\bibitem[Gretler\;and\;Regenfelder(2008)]{gret} W. Gretler, R. Regenfelder,  Effects of radiative transfer on strong shock waves of variable energy propagating in a dusty gas, Phys. Scripta 77 (2008).

\bibitem[Laumbach\;and\;Probstein(1970)]{laumbach} D.D. Laumbach, R.F. Probstein, A Point Explosion in a Cold Exponential Atmosphere: Part 2. Radiating Flow, Journal of Fluid Mechanics, J. Fluid Mech. 40 (1970) 833-858.
 
\bibitem[Rosenau\;and\;Frankenthal(1976)]{rosenau} P. Rosenau, S. Frankenthal, Equatorial propagation of axisymmetric magnetohydrodynamic shocks, Phys. Fluids 19 (1976) 1889-1899.



\bibitem[Vishwakarma \;et\; al.(2007)]{vish8} J.P. Vishwakarma, V. Chaube and A. Patel, Self-similar Solution of a Shock Propagation in a Non-ideal Gas, Int. J. Appl. Mech. Eng. 12 (2007) 813-829.

\bibitem[Zel'dovich\;and\;Raizer(1967)]{zel} YA. B. Zel'dovich and YU. P. Raizer, Physics of Shock Waves and High Temperature Hydrodynamic Phenomena, Vol. II., Academic Press, New York, 1967.

\bibitem[[Vishwakarma\;and\;Nath(2009)]{na09} J.P. Vishwakarma, G. Nath, A self- similar solution of a shock propagation in a mixture of a non-ideal gas and small solid particles, Meccanica, 44 (2009) 239.

\bibitem[Moelwyn-Hughes(1961)]{moel} E.A. Moelwyn-Hughes, physical chemistry, pergamon press, London ,1961.



\end{thebibliography}
\end{document}